\documentclass[aps,prl,twocolumn,superscriptaddress,showpacs,floatfix]{revtex4}

\usepackage{mathrsfs}
\usepackage{graphicx}

\begin{document}

\title{Classical Nature of the Evolution of Dark Energy Density}

\author{Wei Yuan}
\affiliation{Department of Physics, Peking University, Beijing
100871, China}

\author{Yu-xin Liu}
\email[corresponding author, ]{yxliu@pku.edu.cn}
\affiliation{Department of Physics, Peking University, Beijing
100871, China} \affiliation{The Key Laboratory of Heavy Ion Physics,
Ministry of Education,Beijing 100871, China } \affiliation{Center of
Theoretical Nuclear Physics, National Laboratory of Heavy Ion
Accelerator, Lanzhou 730000, China}

\date{\today}

\begin{abstract}
By ignoring the local density fluctuations, we construct an uniform
Higgs-field's (inflaton's) quantum theory
with varying effective Planck constant ($\hbar_{v}(t) \propto
R(t)^{-3}$) for the evolution of the dark energy density during the
epoch after inflation. With presumable sufficient inflation in the
very early period (time-scale is $t_{inf}$), so that
$\hbar_{v}\rightarrow 0$, the state of universe decomposes into some
decoherent components, which could be the essential meaning of phase
transition, and each of them could be well described by classical
mechanics for an inharmonic oscillator in the corresponding
potential-well with a viscous force.
We find that the cosmological constant at present is
$\Lambda_{now}\approx2.05\times 10^{-3}$~eV, which is almost
independent of the choice of potential for inflaton, and agrees
excellently with the recent observations.
%
In addition, we find that, during the cosmic epoch after inflation,
the dark energy is almost conserved as well as the matter's energy,
therefore the ``why now" problem can be avoided.
\end{abstract}


\maketitle

\newpage

\parindent=20pt
{\it Introduction.--}Up to now, the dominant energy density which
governs the evolution of the whole universe is still dark energy
density $\Omega$ \cite{dominant}, which can also be understood in
terms of Einstein's cosmological constant $\Lambda=\Omega^{1/4}$.
However, the size of a small mass scale, $\Lambda\sim 10^{-3}$~eV,
has not yet been derived from a fundamental theory, and its nature
has not been understood either. On the other hand, in very early
universe, the dark energy density is expected to maintain at
extremely high level for a while to realize the well known
inflation~\cite{Guth-1}, finally, it rolls down the hill of the
potential during the epoch after inflation \cite{Linde}. If one
expects to understand the evolution of dark energy density and the
accelerative expansion of universe during epoch after the abrupt
inflation, a detailed theory for  the rolling-down process (perhaps
for the inflation process itself) seems to be required. Such a
theory should also avoid the familiar ``why  now" problem: why do we
find ourself in such a epoch when the cosmological constant is near
zero~\cite{why-1} and why do we live during an era  when the energy
densities in matter and dark energy are comparable~\cite{why-2}.

In this paper, we establish a quantum theory of uniform scalar field
for the evolution of dark energy, where the local density
fluctuations are ignored, and it is expected to carry the leading
order effects of the evolution of the dark energy density. In
addition, our calculations will make no use of the conceptions such
as, effective potential, statistical ensembles, finite temperature
quantum field theory (QFT), which are defined to describe a static,
equilibrium system, however, as we will see below, the real
situation is non-equilibrium. Nevertheless, as we give up here the
concept of statistical ensembles, it seems to rise another familiar
problem: in a pure quantum picture, how can we define the events
such as phase transition or spontaneous symmetry breaking? In
traditional quantum measurement theory, events are related to the
entanglement between the apparatus and the system which attracts our
interests. Thus, we could not imagine any event that has emerged in
a pure quantum evolution before we take an apparatus to observe it,
because such imaginations will destroy the coherence between each
component of the quantum state. We thought it is trustless to
imagine the existence of some environments which will induce the
event of phase transition to occur in the early universe. The
crucial problem turns out to be whether we can find a way to realize
the decoherece between each phase in a pure quantum process without
any environment. To this problem, the irradiative arguments have
been put forward (see for example Ref.~\cite{textbook}). In this
paper, we propose that the inflation of early universe will indeed
help us to realize the events of phase transition and the
spontaneous symmetry breaking. Moreover, the followed evolution of
dark energy density can be well handled by a classical theory.

{\it Formalism.--}We start with the simplest model for a single
scalar Higgs field $\phi$. However, as we will see below, the
results are almost independent of the choice of potential $V(\phi)$
and can be easily generalized to the case with multiple Higgs
fields. By excluding any local density fluctuation, the
space-differential term can be deleted from the Lagrangian
\begin{equation}
\mathscr{L}(\phi)=\frac{1}{2}(\frac{d\phi}{dt})^{2}
+\frac{1}{2}\mu^{2}\phi^{2}-\frac{1}{4}\lambda\phi^{4}
-\frac{\mu^{4}}{4\lambda},
\end{equation}
and the path integral formalism is
\begin{equation}
\int{[D\phi(t)}]e^{i\int{d^{4}x\mathscr{L}}}=\int{[D\phi(t)]
e^{i\int{dt} \hbar_{v}(t)^{-1}\mathscr{L}(\phi,\frac{d\phi}{dt})}},
\end{equation}
with $\hbar_{v}\equiv v^{-1}$ and $v$ being the volume of the
universe. Because the universe is expanding with respect to time,
the effective Planck constant $\hbar_{v}$ is time dependent.
However, the system can still be quantized by the canonical method
which is equivalent to the path integral approach. With such a
canonical quantization, we obtain a Shr\"{o}dinger-like equation
with varying Planck constant
\begin{equation}
i\hbar_{v}(t)\frac{\partial}{\partial
t}\Psi(\phi,t)=[-\frac{\hbar_{v}(t)^{2}}{2}\frac{\partial^{2}}{\partial
\phi^{2}}+V(\phi)]\Psi(\phi,t).
\end{equation}
It should be noted that the mass dimension of $\hbar_{v}$ is 3, thus
the mass dimension of Hamiltonian $\mathscr{H}$ is 4, then
$\mathscr{H}$ represents actually the operator of energy density
rather than energy. Since we only take the uniform scalar field
configurations into account, the expectation value
$\Omega(t)\equiv\langle\Psi(t)| \mathscr{H}|\Psi(t)\rangle$ is just
the dark energy density. Considering the Hermite property of the
Hamiltonian, we have the time-differential of $\Omega$ as
\begin{equation}
\frac{d\Omega(t)}{dt}=2\hbar_{v}^{-1}\frac{d\hbar_{v}}{dt}T(t),
\end{equation}
where
$T=\langle\Psi(t)|-\frac{\hbar_{v}(t)^{2}}{2}\frac{\partial^{2}}{\partial
\phi^{2}}|\Psi(t)\rangle$ is the average kinetic energy density. It
has been well established that, after sufficient inflation, the
space-time can be described by the de-Sitter metric
\begin{equation}
ds^{2}=dt^{2}-R(t)^{2}d\vec{x}^{2},
\end{equation}
where the first order Friedmann equation for $R(t)$ reads
\begin{equation}
R^{-1}\frac{dR}{dt}=(\frac{8\pi}{3} G \Omega)^{1/2},
\end{equation}
where $G$ is the gravitation constant. Considering the relation
$\hbar_{v}=v^{-1} \propto R^{-3}$, we can rewrite the first order
Friedmann equation as
\begin{equation}
\hbar_{v}^{-1}\frac{d \hbar_{v}}{dt}=-(24\pi G \Omega)^{1/2}.
\end{equation}
Substituting Eq.~(7) into Eq.~(4), we get
\begin{equation}
\frac{d\Omega(t)}{dt}=-2(24\pi G \Omega)^{1/2}T(t).
\end{equation}
On one hand, Eqs.~(3) and (7) could describe a theoretically
solvable quantum system. However, they are worthy only if the
quantum fluctuations are indispensable, for instance, during the
cosmic epoch of inflation. One can imagine that $\hbar_{v}$ is very
large in that period, it forces us to make use of Eqs.~(3) and (7)
exactly (we have not yet investigated such process, inflation
itself. Moreover, in such an epoch, de-Sitter metric is plausible).
On the other hand, once inflation has lasted continuously for a
while ($t_{inf}$) and quenched, Eqs.~(7) and (8) imply that the
effective Planck constant becomes very small and the energy density
decreases to a level well below the potential density barrier. At
that moment, one might have reasons to expect that the
quantum-mechanical tunneling effects are strongly restrained by the
smallness of $\hbar_{v}$ and $\Omega$, thus the wave function in the
negative $\phi$ region will decohere with the wave function in the
positive $\phi$ region. We emphasize that the decoherence of these
two equivalent components is a reasonable signal of the event of
phase transition (viz. spontaneous symmetry breaking) which occurs
at the time $t_{c}\simeq t_{inf}$. The followed evolution of the
localized wave packet is known as rolling-down. Guth had
investigated a similar process (with a parameter in Eq.~(1) being
set as $\lambda = 0 $\cite{Guth-2}), and pointed out that the
evolution of such a quantum wave packet can be understood with a
probability distribution which describes the classical trajectories
rolling in the well. Actually, we do not need to worry about the
problem of which trajectory we should choose, because these
trajectories are different from each other only by a series of
time-translations, and the typical translation's scale is extremely
small compared to nowadays cosmic age. The problem is essentially
that, after a long time evolution, no observation can distinguish
these different trajectories. Therefore, the evolution of each
localized wave function in the corresponding well of the potential
density surface can be well described by classical mechanics. Here,
it means that, during the cosmic epoch after inflation, we can take
the classical kinematics as the complement for Eq.~(8), so that it
is solvable. The dark energy density in each well can then be given
as
\begin{equation}
\Omega=T+V(\phi)=\frac{1}{2}(\frac{d\phi}{dt})^{2}+V(\phi).
\end{equation}
Combining Eq.~(9) with Eq.~(8), one can easily obtain a solvable
classical equation for $\phi(t)$
\begin{equation}
\frac{d^{2}\phi}{dt^{2}}=-\frac{dV(\phi)}{d\phi}-(24\pi G
\Omega)^{1/2}\frac{d\phi}{dt}.
\end{equation}

Eq.~(10) is just the equation of the motion of a point mass evolving
in the potential $V(\phi)$ in the presence of a viscous force $-(24
\pi G \Omega)^{1/2} \frac{d \phi}{d t}$. Such a viscous force
results naturally from the quantum effects and the gravity effects.
It should be emphasized that Eq.~(10) can be independently derived
by combining the full Friedmann equations (the first order and the
second order) with the classical mechanics of uniform Higgs field,
which has been implied in the chaotic inflation theory
\cite{Linde-2}. However, our derivation for Eq.~(10), which has not
made use of the second order Friedmann equation, indicates that the
Shr\"{o}dinger-like equation in Eq.~(3) is not only naturally
consistent with the second order Friedmann equation but also a
proper substitution for Eq.~(10) during the epoch of inflation.
Here, we just need to note that Eq. (10) can only be used precisely
for the epoch after inflation but it is not true for the inflation
process itself.

Since the $\phi$ is expected to be near the bottom of the potential
density surface $V(\phi)$ (here we choose
$\phi\sim\frac{\mu}{\lambda^{1/2}}$), we can then, at the lowest
order, approximate the potential $V(\phi)$ to be
\begin{equation}
V(\phi_{f})\approx \mu^{2} \phi_{f}^{2},
\end{equation}
with $\phi=\frac{\mu}{\lambda^{1/2}}+\phi_{f}$, and the
corresponding equation for $\phi_{f}(t)$ becomes
\begin{equation}
\frac{d^{2}\phi_{f}}{dt^{2}}=-\omega^{2}\phi_{f}-(24\pi G )^{1/2}
[\frac{1}{2}(\frac{d\phi_{f}}{dt})^{2}+\frac{1}{2}\omega^{2}
\phi_{f}^{2}]^{1/2}\frac{d\phi_{f}}{dt}\, ,
\end{equation}
with $\omega=\sqrt{2}\mu$ (we will verify the validity of this
approximation later). It is obvious that there still exist
difficulties in solving Eq.~(12) exactly. Nevertheless, there is a
reasonable way to obtain the evolution of $\Omega(t)$. If we ignore
the viscous force at first, the system reduces to a harmonic
oscillator, its time-period is $\frac{2\pi}{\omega}$. Then we turn
on the viscous force, but do not change the behavior of the harmonic
oscillation in one time-period. After some calculations, we obtain
the negative work made by the viscous force in this period as
\begin{equation}
W_{T}=-\frac{2\pi}{\omega}(24\pi G)^{1/2}\Omega^{3/2}.
\end{equation}
Since the ratio $W_{T}/\Omega = - \frac{2 \pi}{\omega} (24 \pi G
\Omega)^{1/2} \propto \frac{\Omega^{1/2}}{M_{p}\mu}$, where $M_{p} =
G^{-1/2}$ is the Planck scale of energy, is expected to be very tiny
and the present observations indicate that such an expectation is
true, we can really perform such a calculation in one period without
changing the harmonic oscillation. Furthermore, during the epoch
after inflation, $\frac{2\pi}{\omega}$ should be a very short time
scale compared to the time scale of $\Omega(t)$'s evolution (as we
will see self-consistently below), we can then take approximations
$W_{T}\rightarrow d\Omega$ and $\frac{2\pi}{\omega}\rightarrow dt$.
It turns out that Eq.~(13) is a representation of the equation for
the evolution of dark energy density, and can be written explicitly
as
\begin{equation}
\frac{d\Omega}{dt}=-(24\pi G)^{1/2}\Omega^{3/2}.
\end{equation}

One can easily understand Eq.~(14) by simply substituting
$T=\Omega/2$ into Eq.~(8), since it is the nature of quickly
harmonic oscillation. Furthermore, it should be emphasized that the
intrinsic parameters $\mu$ and $\lambda$ of the quantum filed theory
do not appear in Eq.~(14), however, their effect is involved in the
initial conditions, because Eq.~(14) is valid only after some time
$t_{0}> t_{inf}$ which is the time for Eq.~(11) to be valid, and
$t_{0}$ depends definitely on $\mu$ and $\lambda$. Nevertheless, we
take the initial conditions as $t_{0}$ and $\Omega_{0}$, with which
we can easily find out the definite solution for Eq.~(14) as
\begin{equation}
\Omega(t)=\frac{M_{p}^{2}}{6\pi [(t-t_{0})+t_{\Omega_{0}}]^{2}},
\end{equation}
with $t_{\Omega_{0}}$ satisfying $\Omega_{0}=\frac{M_{p}^{2}}{6\pi
t_{\Omega_{0}}^{2}}$. On one hand, we need $\Omega_{0}$ to be small
enough to ensure Eq.~(11), which requires $\Omega_{0}\ll
\frac{\mu^{4}}{\lambda}$. On the other hand, we hope
$t_{\Omega_{0}}$ is small enough compared to $t_{now}\simeq
1.37\times 10^{10}$~years~ (the age of universe with new accuracy
from WMAP date) so that we can precisely make use of Eq.~(15) in the
epoch after inflation by simply ignoring $t_{\Omega_{0}}$, which
requires $\Omega_{0}\gg \frac{M_{p}^{2}}{6 \pi t_{now}^{2}}$.
Altogether, it requires
\begin{equation}
\frac{\mu^{4}}{\lambda}\gg \frac{M_{p}^{2}}{6 \pi t_{now}^{2}}\simeq
1.78 \times 10^{-47}\, \mbox{GeV}^{4} \, .
\end{equation}
We are glad to note that any reasonable choice for $\mu$ and
$\lambda$ can perfectly satisfy such a condition. The last problem
is that we have not found out a close formalism for
$t_{0}(\mu,\lambda)$ nor for $t_{inf}(\mu,\lambda)$, it might need a
full quantum treatment on Eqs.~(3) and (7). Fortunately, whatever
$t_{0}(\mu,\lambda)$ should be, it can not change the solid fact
that $t_{now} \gg t_{0}>t_{inf}$. By rewriting $ t_{\Omega_{0}}
-t_{0} \equiv t_{eff}$, we have the varying cosmological constant as
\begin{equation}
\Lambda = \Big[\frac{M_{p}}{\sqrt{6\pi} (t+t_{eff})}\Big]^{1/2} \, .
\end{equation}
When we apply Eq.~(17) to the universe after inflation (actually,
Eq.~(17) can only be used in this case, since for very early
universe, detailed treatment on the quantum nature of inflation is
required, and the de-Sitter metric is plausible), $t_{eff} \ll t$,
we can take $t_{eff}$ as zero. We obtain then the evolution of the
cosmological constant with respect to time as shown in Fig.~1.
\begin{figure}[!htb]
\includegraphics[scale=0.8,bb=0 0 300 250,clip]{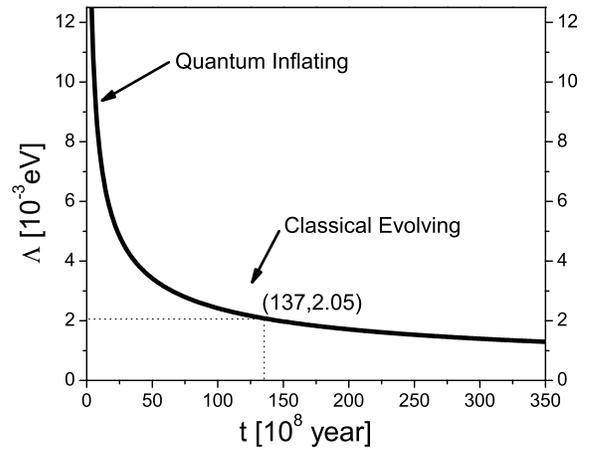}
\caption{\label{1} The calculated evolution of Einstein's
cosmological constant with respect to time (with Planck scale of
energy $M_{p}\simeq1.22\times 10^{19}GeV$). }
\end{figure}
With the age of the present universe $t_{now}\simeq 1.37\times
10^{10}$~years and the Planck scale of energy $M_{p} = G^{-1/2}
\approx 1.22 \times 10^{19}$~GeV, we obtain the dark energy density
or the Einstein's cosmological constant at present as $
\Omega_{now}\simeq 1.78 \times 10^{-47} \mbox{GeV}^{4} $, or
$\Lambda_{now}\simeq 2.05 \times 10^{-3}$~eV. It is apparent that
such a result agrees excellently with the recent SDSS and WMAP
observation~\cite{Tegmark}: $\Lambda=2.14 \pm0.13 \times 10^{-3}$~
eV. As for the behavior shown in Fig.~1 in the quantum inflating
epoch, even though it can not be trusted precisely, it is still
consistent with the inflation picture which has been presumed in our
calculations.

Recalling Eq.~(17) with approximation $t_{eff} = 0$, one can easily
realize that a cosmological constant $\Lambda = 2 \Lambda_{now}$ can
only be observed at $1.03\times 10^{10}$~years earlier, and $\Lambda
=\frac{1}{2} \Lambda_{now}$ can only be observed at $4.11\times
10^{10}$~years later. It means that, in the cosmic epoch after
inflation, the cosmological constant evolves very slowly. Moreover,
by using the identity
\begin{equation}
\frac{d(\Omega/\hbar_{v})}{(\Omega/\hbar_{v})dt}\equiv \Omega^{-1}
\frac{d\Omega}{dt}-\hbar_{v}^{-1}\frac{d\hbar_{v}}{dt} \, ,
\end{equation}
and Eqs.~(7) and (14), we have
\begin{equation}
\frac{d(\Omega v)}{dt}=0.
\end{equation}
It is presumable that, once the inflation has already ended, the
probability of local materialization from false vacuum could be very
small. Therefore, Eq.~(19) means that the total dark energy is
almost conserved just as the total energy included in matter (both
luminous matter and dark matter) does. The above result indicates
that, during the cosmic epoch after inflation, the ratio between the
density of dark energy and the energy density of the matter,
$\frac{\Omega_{DE}}{\Omega_{matt}}$, is almost a constant (present
observation shows $\frac{\Omega_{DE}}{\Omega_{matt}} \approx
\frac{7}{3}$). Altogether, the ``why now" problem is solved.
However, it leaves a problem ``why $\frac{7}{3}$" to be answered in
the inflation epoch.

The presently obtained results indicate more interestingly is that,
after the presumable sufficient inflation, the cosmology constant is
almost independent of the intrinsic parameters of the quantum field
theory. It means that one can choose any $V(\phi)$ which could have
arbitrary minimal positions, however, the inflation causes the
components located in these valleys decohere from each other, and no
matter where we live, the observed cosmological constant is the
same. If there are $N_{\phi}$ independent Higgs fields, which belong
to different interior spaces, the cosmological constant should be
$N_{\phi}^{1/4}\times 2.05 \times 10^{-3}$~eV (note that energy
density is additive quantity, but it is not true for cosmological
constant). If the multiple Higgs fields belong to a single interior
space, it should not change our result for the cosmological constant
($2.05 \times 10^{-3}$~eV). Precise observation for cosmologcial
constant can then be taken as a signature to determine the
$N_{\phi}$ and test our theory itself. Our present results seem to
support the case of $N_{\phi}=1$ (Viz. all possible Higgs fields
should belong to an interior space and hover around some valley of
the potential density surface during the epoch after inflation).

{\it Summary.--} In this paper, we try to show that the relation
between the epoch after inflation and the epoch during inflation is
very subtle. During inflation, the quantum fluctuations are very
strong, as Guth pointed out~\cite{Guth-2} that, one could expect to
understand the production mechanism for essentially all matters,
energy, entropy in such a process.
Nevertheless, whatever the precise inflation picture should be, the
rough properties of it seem to be sufficient to set the initial
conditions for the precise interpretation of the slowly evolving
dark energy density during the epoch after inflation. Thus, without
adjusting any parameter, we give an evolution behavior of the dark
energy density or the Einstein's cosmological constant, and the
results are agree excellently with the recent SDSS and WMAP
observations. In addition, the ``why now" problem is solved in our
present approach. Thus we suppose that the present evolving
cosmological constant is governed by a classical inharmonic
oscillation and consistent with the inflation picture, but it is
irrelevant to the details of quantum inflation itself. It provides
then a classical nature for the evolution of the dark energy density
after inflation.
\bigskip

\begin{acknowledgments}
This work was supported by the National Natural Science Foundation
of China (NSFC) under contract Nos. 10425521 and 10575004, the
Major State Basic Research Development Program under contract No.
G2000077400, the research foundation of the Ministry of Education,
China (MOEC), under contact No. 305001 and the Research Fund for
the Doctoral Program of Higher Education of China under grant No
20040001010. One of the authors (YXL) thanks also the support of
the Foundation for University Key Teacher by the MOEC.

\end{acknowledgments}

\end{document}